\documentclass[12pt]{article}
\usepackage{graphicx}
\usepackage{color}
\usepackage{authblk}
\def\hybrid{\topmargin 0pt      \oddsidemargin 0pt
        \headheight 0pt \headsep 0pt
       \voffset-1cm
        \textwidth 6.25in       % A4 paper
       \textheight 9.5in       % A4 paper
        \marginparwidth 0.0in
        \parskip 5pt plus 1pt   \jot = 1.5ex}
\catcode`\@=11
\def\marginnote#1{}

\newcount\hour
\newcount\minute
\newtoks\amorpm
\hour=\time\divide\hour by60
\minute=\time{\multiply\hour by60 \global\advance\minute by-\hour}
\edef\standardtime{{\ifnum\hour<12 \global\amorpm={am}%
        \else\global\amorpm={pm}\advance\hour by-12 \fi
        \ifnum\hour=0 \hour=12 \fi
        \number\hour:\ifnum\minute<10 0\fi\number\minute\the\amorpm}}
\edef\militarytime{\number\hour:\ifnum\minute<10 0\fi\number\minute}

\def\draftlabel#1{{\@bsphack\if@filesw {\let\thepage\relax
   \xdef\@gtempa{\write\@auxout{\string
      \newlabel{#1}{{\@currentlabel}{\thepage}}}}}\@gtempa
   \if@nobreak \ifvmode\nobreak\fi\fi\fi\@esphack}
        \gdef\@eqnlabel{#1}}
\def\@eqnlabel{}
\def\@vacuum{}
\def\draftmarginnote#1{\marginpar{\raggedright\scriptsize\tt#1}}

\def\draftlabel#1{{\@bsphack\if@filesw {\let\thepage\relax
   \xdef\@gtempa{\write\@auxout{\string
      \newlabel{#1}{{\@currentlabel}{\thepage}}}}}\@gtempa
   \if@nobreak \ifvmode\nobreak\fi\fi\fi\@esphack}
        \gdef\@eqnlabel{#1}}
\def\@eqnlabel{}
\def\@vacuum{}
\def\draftmarginnote#1{\marginpar{\raggedright\scriptsize\tt#1}}

\def\draft{\oddsidemargin -.5truein
        \def\@oddfoot{\sl preliminary draft \hfil
        \rm\thepage\hfil\sl\today\quad\militarytime}
        \let\@evenfoot\@oddfoot \overfullrule 3pt
        \let\label=\draftlabel
        \let\marginnote=\draftmarginnote
   \def\@eqnnum{(\theequation)\rlap{\kern\marginparsep\tt\@eqnlabel}%
\global\let\@eqnlabel\@vacuum}  }

\def\numberbysection{\@addtoreset{equation}{section}
        \def\theequation{\thesection.\arabic{equation}}}

\def\underline#1{\relax\ifmmode\@@underline#1\else
        $\@@underline{\hbox{#1}}$\relax\fi}

\def\titlepage{\@restonecolfalse\if@twocolumn\@restonecoltrue\onecolumn
     \else \newpage \fi \thispagestyle{empty}\c@page\z@
        \def\thefootnote{\fnsymbol{footnote}} }

\def\endtitlepage{\if@restonecol\twocolumn \else  \fi
        \def\thefootnote{\arabic{footnote}}
        \setcounter{footnote}{0}}  %\c@footnote\z@ }
\relax

\numberbysection
\hybrid

\newfont{\Bbb}{msbm10 scaled 1\@ptsize00}
\newfont{\Bbbb}{msbm7 scaled 1\@ptsize00}

\newcommand{\DDD}{\raise-1pt\hbox{$\mbox{\Bbbb D}$}}

\newcommand{\UUU}{\raise-1pt\hbox{$\mbox{\Bbbb U}$}}

\newcommand{\z}{\raise-1pt\hbox{$\mbox{\Bbbb Z}$}}

\def\beq{\begin{equation}}
\def\eeq{\end{equation}}
\def\p{\partial}

\begin{document}

\begin{titlepage}

\title{Tau-function of the multi-component CKP hierarchy}

\author{A. Zabrodin\thanks{Skolkovo Institute of Science and
Technology, 143026, Moscow, Russia and
National Research University Higher School of
Economics,
20 Myasnitskaya Ulitsa, Moscow 101000, Russia, 
and NRC ``Kurchatov institute'', Moscow, Russia;
e-mail: zabrodin@itep.ru}}

\date{August 2023}
\maketitle

\vspace{-7cm} \centerline{ \hfill ITEP-TH-18/23}\vspace{7cm}

\begin{abstract}

We consider multi-component Kadomtsev-Petviashvili hierarchy of type C 
(the multi-component CKP hierarchy) originally defined with the help
of matrix pseudo-differential operators via the Lax-Sato formalism. 
Starting from the bilinear relation for the wave functions, 
we prove existence of the tau-function for the multi-component 
CKP hierarchy and provide a formula which expresses the wave functions
through the tau-function. We also find
how this tau-function is related to the tau-function of the 
multi-component Kadomtsev-Petviashvili hierarchy. 
The tau-function of the multi-component
CKP hierarchy satisfies an integral relation which, unlike the 
integral relation for the latter tau-function, is no longer bilinear
but has a more complicated form.

\end{abstract}

\end{titlepage}

\noindent
Key words: integrable hierarchies, tau-function, multi-component
CKP hierarchy, wave functions

\tableofcontents

\section{Introduction}

Integrable nonlinear partial differential equations are known to
form infinite hierarchies of consistent equations. One of the most 
widely known is the Kadomtsev-Petviashvili (KP) hierarchy \cite{DJKM83}.
The set of independent variables is ${\bf t}=\{t_1, t_2, t_3, \ldots \}$
(``times'') which are in general complex numbers.
In the Lax-Sato formalism, the main object 
is the Lax operator which is a pseudo-differential operator of the form
\beq\label{int1}
{\cal L}=\p_x +u_1\p_x^{-1}+u_2 \p_x^{-2}+\ldots
\eeq
where the coefficients $u_i$ are functions of $x, {\bf t}$. The KP
hierarchy is the set of evolution equations for $u_i$'s written in the 
Lax form as
\beq\label{int2}
\p_{t_{k}}{\cal L}=[B_{k}, \, {\cal L}], 
\quad B_{k} = \Bigl ({\cal L}^k\Bigr )_+, 
\quad k=1,2,3, \ldots ,
\eeq
where $(\ldots )_+$ means the differential part of a pseudo-differential operator,
i.e. the sum of all terms with $\p_x^k$, where $k\geq 0$. In particular,
$B_1=\p_x$, so
$\p_{t_1}{\cal L}=\p_x {\cal L}$ which means that we can identify
$t_1$ with $x+\mbox{const}$. 

The general solution of the KP hierarchy 
is given by the tau-function $\tau =\tau (x, {\bf t})$,
the functions $u_i$ being given by 
certain derivatives of $\log \tau$. The tau-function satisfies an
infinite set of bilinear equations of the Hirota type which can be 
encoded in one generating integral bilinear relation \cite{DJKM83}.

The CKP hierarchy \cite{DJKM81a}--\cite{KZ21} 
is obtained from the KP one by imposing the constraint
\beq\label{int3}
{\cal L}^{\dag}=-{\cal L}.
\eeq
Here $\dag$ means the formal adjoint operator
defined by the rule $\Bigl (f(x)\circ 
\p_x^{n}\Bigr )^{\dag}=(-\p_x)^n \circ f(x)$. This constraint is
preserved by the flows (\ref{int2}) with odd $k$ and is destroyed 
by the flows with even $k$. Therefore, in order to define the 
CKP hierarchy one should restrict the set of independent variables
to be ${\bf t}^{\rm o}=\{ t_1, t_3, t_5, \ldots \}$ putting
$t_{2k}=0$ for all $k$. 

The multi-component (matrix) extension of the KP hierarchy was
introduced in \cite{DJKM81}, see also \cite{KL93,TT07}. In this case
the coefficients $u_i$ in (\ref{int1}) are $n \times n$ matrices,
and the set of independent variables is extended by additional 
integrable flows. The tau-function also becomes an $n\times n$
matrix. In this paper we introduce a subhierarchy of the multi-component
KP hierarchy which can be naturally 
regarded as a multi-component extension
of the CKP hierarchy. We also define its tau-function and show how
it is related to the tau-function of the multi-component KP hierarchy.
The multi-component CKP hierarchy was also discussed
in the recent paper \cite{KL23}.

The paper is organized as follows. In section 2 we remind the reader
the main facts on the multi-component KP hierarchy. We present both
Lax-Sato and bilinear formalism. Section 3 is devoted to the 
multi-component CKP hierarchy. We start from the Lax-Sato formalism
and then prove the existence of the matrix 
tau-function $\tau^{\rm CKP}$. Section 4 contains some concluding
remarks.

\section{The multi-component KP hierarchy}

Here we briefly review the main facts about the multi-component 
KP hierarchies following \cite{TT07,Teo11}.

\subsection{Lax-Sato formalism for the multi-component KP hierarchy}

The independent variables are $n$ infinite sets of continuous ``times''
$$
{\bf t}=\{{\bf t}_1, {\bf t}_2, \ldots , {\bf t}_n\}, \qquad
{\bf t}_{\alpha}=\{t_{\alpha , 1}, t_{\alpha , 2}, t_{\alpha , 3}, \ldots \, \},
\qquad \alpha = 1, \ldots , n.
$$
It is convenient to introduce also the variable $x$ such that 
\beq\label{multi2}
\p_x =\sum_{\alpha =1}^n \p_{t_{\alpha , 1}}.
\eeq
The hierarchy is an infinite set of evolution equations in the times 
${\bf t}$ for matrix
functions of the variable $x$. 

In the Lax-Sato formalism, the main object 
is the Lax operator which is a pseudo-differential operator of the form
\beq\label{multi3}
{\cal L}=\p_x +u_1\p_x^{-1}+u_2 \p_x^{-2}+\ldots
\eeq
where the coefficients $u_i=u_i(x, {\bf t})$ are $n\! \times \! n$ matrices.
The coefficient functions $u_k$ depend on $x$ and also on all the times:
$$
u_k(x, {\bf t})=u_k(x+t_{1,1}, x+t_{2,1}, \ldots , x+t_{n,1};
t_{1,2}, \ldots , t_{n,2}; \ldots ).
$$ 
Besides, there are
other matrix pseudo-differential operators ${\cal R}_1, \ldots , {\cal R}_n$ 
of the form
\beq\label{multi5}
{\cal R}_{\alpha}=E_{\alpha}+ u_{\alpha , 1}\p_x^{-1}+u_{\alpha , 2}\p_x^{-2}+\ldots ,
\eeq
where $E_{\alpha}$ is the $n \! \times \! n$ matrix with the 
$(\alpha , \alpha )$ element
equal to 1 and all other components equal to $0$, and 
$u_{\alpha , i}$ are also
$n \! \times \! n$ matrices. 
The operators ${\cal L}$, ${\cal R}_1, \ldots , {\cal R}_n$
satisfy the conditions 
\beq\label{multi6}
{\cal L}{\cal R}_{\alpha}={\cal R}_{\alpha}{\cal L}, \quad
{\cal R}_{\alpha}{\cal R}_{\beta}=\delta_{\alpha \beta}{\cal R}_{\alpha}, \quad
\sum_{\alpha =1}^n {\cal R}_{\alpha}=I,
\eeq
where $I$ is the unity matrix.
The Lax equations of the hierarchy which define evolution in the times read
\beq\label{multi8}
\p_{t_{\alpha , k}}{\cal L}=[B_{\alpha , k}, \, {\cal L}], \quad
\p_{t_{\alpha , k}}{\cal R}_{\beta}=[B_{\alpha , k}, \, {\cal R}_{\beta}],
\quad B_{\alpha , k} = \Bigl ({\cal L}^k{\cal R}_{\alpha}\Bigr )_+, 
\quad k=1,2,3, \ldots ,
\eeq
where $(\ldots )_+$ means the differential part of a pseudo-differential operator,
i.e. the sum of all terms with $\p_x^k$, where $k\geq 0$. 

Let us introduce the matrix pseudo-differential 
``wave operator'' ${\cal W}$ with matrix elements
\beq\label{m113}
{\cal W}_{\alpha \beta} = \delta_{\alpha \beta}+\sum_{k\geq 1}
\xi_{k,\alpha \beta}(x,{\bf t})\p_x^{-k},
\eeq
where $\xi_{k,\alpha \beta}(x,{\bf t})$ are 
some matrix functions. The operators ${\cal L}$ and ${\cal R}_{\alpha}$ are obtained 
from the ``bare'' operators $I\p_x $ and $E_{\alpha}$ by ``dressing'' 
by means of the wave operator:
\beq\label{multi12}
{\cal L}={\cal W}\p_x {\cal W}^{-1}, \qquad
{\cal R}_{\alpha}={\cal W}E_{\alpha} {\cal W}^{-1}.
\eeq
Clearly, there is an ambiguity in the definition of the dressing operator: it can be multiplied
from the right by any pseudo-differential operator with constant coefficients 
commuting with $E_{\alpha}$ for any $\alpha$.

A very important role in the theory is played by the 
wave function $\Psi$ and its dual
$\Psi^{*}$.
The wave function 
is defined as a result of action of the wave operator 
to the exponential function:
\beq\label{m113a}
\Psi (x,{\bf t}; z)={\cal W}
\exp \Bigl (xzI+\sum_{\alpha =1}^n E_{\alpha}\xi ({\bf t}_{\alpha}, z)\Bigr ),
\eeq
where we use the standard notation
$$
\xi ({\bf t}_{\alpha}, z)=\sum_{k\geq 1}t_{\alpha , k}z^k.
$$
By definition, the operators $\p_x^{-k}$ with negative powers act 
to the exponential function as $\p_x^{-k}e^{xz}=z^{-k}e^{xz}$. 
The wave function depends on the spectral parameter $z$ 
which does not enter the auxiliary
linear problems explicitly. 
The dual wave function is introduced by the formula
\beq\label{m113b}
\Psi ^{*} (x,{\bf t}; z)=\exp \Bigl (-xzI-\sum_{\alpha =1}^n 
E_{\alpha}\xi ({\bf t}_{\alpha}, z)\Bigr )
{\cal W}^{-1}.
\eeq
Here we use the convention that 
the operators $\p_x$ 
which enter ${\cal W}^{-1}$ act to the left rather than to the right, 
the left action being defined as 
$f\stackrel{\leftarrow}{\p_x} \equiv -\p_x f$. 
Clearly, the expansion of the wave function
as $z\to \infty$ is as follows:
\beq\label{multi13}
\Psi_{\alpha \beta} (x,{\bf t}; z)=e^{xz+\xi ({\bf t}_{\beta}, z)}
\Bigl (\delta_{\alpha \beta}+\xi^{(1)}_{\alpha \beta}z^{-1}+
\xi^{(2)}_{\alpha \beta}z^{-2}+\ldots \Bigr ).
\eeq

As is proved in \cite{Teo11}, the wave function 
satisfies the linear
equations
\beq\label{m13c}
\p_{t_{\alpha , m}}\Psi (x,{\bf t}; z)=B_{\alpha ,m} \Psi (x,{\bf t}; z), 
\eeq
where $B_{\alpha ,m}$ is the differential operator (\ref{multi8}), i.e.
$
B_{\alpha ,m}= \Bigl ({\cal W} E_{\alpha}\p_x^m {\cal W}^{-1}\Bigr )_+
$
and the dual wave function satisfies the transposed equations
\beq\label{m13e}
-\p_{t_{\alpha , m}}\Psi^{*} (x,{\bf t}; z)=
\Psi^{*} (x,{\bf t}; z) B_{\alpha ,m} .
\eeq
Again, the operator
$B_{\alpha , m}$ here acts to the left.
In particular, it follows from
(\ref{m13c}), (\ref{m13e}) at $m=1$ that
\beq\label{m13d}
\sum_{\alpha =1}^{n}\p_{t_{\alpha , 1}}
\Psi (x,{\bf t}; z)=\p_x \Psi (x,{\bf t}; z),
\qquad
\sum_{\alpha =1}^{n}\p_{t_{\alpha , 1}}\Psi^{*} (x,{\bf t}; z)=
\p_x \Psi^{*} (x,{\bf t}; z),
\eeq
so the vector field $\p_x$ can be naturally identified with the vector field
$\displaystyle{\sum_{\alpha }\p_{t_{\alpha , 1}}}$.

\subsection{The tau-function}

In this section we put $x=0$ without loss of generality and drop
the letter $x$ from the arguments of functions.

The linear problems (\ref{m13c}, (\ref{m13e}) can be encoded 
in the following bilinear relation for the wave functions:
\beq\label{m3}
\oint_{C_{\infty}}\! dz \, \Psi ({\bf t};z)\Psi^{*} ({\bf t}';z)=0
\eeq
valid for all times ${\bf t}$, ${\bf t}'$.
The integration contour $C_{\infty}$ is a big circle around $\infty$.

We are going to prove that the bilinear relation (\ref{m3}) implies the 
existence of the matrix function $\tau_{\alpha \beta}({\bf t})$ such that
diagonal elements $\tau_{\alpha \alpha}$ are all equal to one and the same function
$\tau =\tau ({\bf t})$ and the 
wave functions
are expressed through $\tau_{\alpha \beta}$ as follows:
\beq\label{m2}
\begin{array}{l}
\displaystyle{\Psi_{\alpha \beta}({\bf t};z)=
\epsilon_{\alpha \beta}\,
\frac{\tau_{\alpha \beta} \left (
{\bf t}-[z^{-1}]_{\beta}\right )}{\tau ({\bf t})}\,
z^{\delta_{\alpha \beta}-1}e^{\xi ({\bf t}_{\beta}, z)},
}
\\ \\
\displaystyle{\Psi_{\alpha \beta}^{*}({\bf t};z)=
\epsilon_{\beta \alpha}\,
\frac{\tau _{\alpha \beta}\left (
{\bf t}+[z^{-1}]_{\alpha}\right )}{\tau ({\bf t})}\,
z^{\delta_{\alpha \beta}-1}e^{-\xi ({\bf t}_{\alpha}, z)}
}.
\end{array}
\eeq
Here $\epsilon_{\alpha \beta}$ is a sign factor:
$\epsilon_{\alpha \beta}=1$ if $\alpha \leq \beta$, 
$\epsilon_{\alpha \beta}=-1$
if $\alpha >\beta$ and we use the standard notation
$$
\left ({\bf t}\pm [z^{-1}]_{\gamma}\right )_{\alpha ,k}=t_{\alpha , k}\pm
\delta_{\alpha \gamma} \frac{z^{-k}}{k}.
$$
The function $\tau_{\alpha \beta}$ 
is called the tau-function. It is a universal dependent variable of the hierarchy.
All other objects (such as
the coefficient functions $u_i$ of the Lax operator) can be 
expressed in terms of it. 
In the bilinear formalism, the $n$-component KP hierarchy 
is the infinite set of bilinear equations
for the tau-function which are encoded in the basic 
relation
\beq\label{m5}
\sum_{\nu =1}^n \epsilon_{\alpha \nu}\epsilon_{\beta \nu}
\oint_{C_{\infty}}\! dz \, 
z^{\delta_{\alpha \nu}+\delta_{\beta \nu}-2}
e^{\xi ({\bf t}_{\nu}-{\bf t}_{\nu}', \, z)}
\tau _{\alpha \nu} \left ({\bf t}-[z^{-1}]_{\nu}\right )
\tau _{\nu \beta}\left ({\bf t}'+[z^{-1}]_{\nu}\right )=0
\eeq 
valid for all ${\bf t}$, ${\bf t}'$. It is obtained by substituting (\ref{m2}) into 
(\ref{m3}).

The proof, first given in \cite{D97}, is based on the bilinear relation (\ref{m3}). Here we present this proof with some modifications. 
Let us represent the wave functions in the form
\beq\label{m6}
\begin{array}{l}
\displaystyle{\Psi_{\alpha \beta}({\bf t};z)=
\epsilon_{\alpha \beta}\,e^{\xi ({\bf t}_{\beta}, z)}
z^{\delta_{\alpha \beta}-1}w_{\alpha \beta}({\bf t}, z)},
\\ \\
\displaystyle{\Psi^{*}_{\alpha \beta}({\bf t};z)=
\epsilon_{\beta \alpha}\,e^{-\xi ({\bf t}_{\alpha}, z)}
z^{\delta_{\alpha \beta}-1}\bar w_{\alpha \beta}({\bf t}, z)}
\end{array}
\eeq
with some matrix functions $w({\bf t}, z)$, $\bar w({\bf t}, z)$. 
Then we can write the bilinear relation as
\beq\label{m7}
\sum_{\gamma =1}^n \epsilon_{\alpha \gamma}\epsilon_{\beta \gamma}
\oint_{C_{\infty}}\! dz \, 
e^{\xi ({\bf t}_{\gamma}-{\bf t}'_{\gamma}, z)}
z^{\delta_{\alpha \gamma}+\delta_{\beta \gamma}-2}
w_{\alpha \gamma}({\bf t}, z)\bar w_{\gamma \beta}({\bf t}', z)=0.
\eeq
In the calculations below, we need the values
\beq\label{m8}
w_{\alpha \beta}({\bf t}, \infty )=:w_{\alpha \beta}({\bf t}),
\quad
\bar w_{\alpha \beta}({\bf t}, \infty )=:\bar w_{\alpha \beta}({\bf t}).
\eeq
We can always normalize them as
\beq\label{m9}
w_{\alpha \alpha}({\bf t})=\bar w_{\alpha \alpha}({\bf t})=1.
\eeq

Setting ${\bf t}-{\bf t}'=[a^{-1}]_{\mu}$ with $\mu =\alpha$ or
$\mu =\beta$, we have from (\ref{m7}) for $\alpha \neq \beta$:
\beq\label{m10}
\begin{array}{c}
\displaystyle{\oint_{C_{\infty}}dz z^{-1}\frac{a}{a-z}\,
w_{\alpha \alpha}({\bf t}, z)\bar w_{\alpha \beta}({\bf t}-
[a^{-1}]_{\alpha}, z)}
\\ \\
\displaystyle{=\oint_{C_{\infty}}dz z^{-1}
w_{\alpha \beta}({\bf t}, z)\bar w_{\beta \beta}({\bf t}-
[a^{-1}]_{\alpha}, z)},
\end{array}
\eeq
\beq\label{m10a}
\begin{array}{c}
\displaystyle{\oint_{C_{\infty}}dz z^{-1}\frac{a}{a-z}\,
w_{\alpha \beta}({\bf t}, z)\bar w_{\beta \beta}({\bf t}-
[a^{-1}]_{\beta}, z)}
\\ \\
\displaystyle{=\oint_{C_{\infty}}dz z^{-1}
w_{\alpha \alpha}({\bf t}, z)\bar w_{\alpha \beta}({\bf t}-
[a^{-1}]_{\beta}, z)}
\end{array}
\eeq
and
\beq\label{m11}
\oint_{C_{\infty}}dz \frac{a}{a-z}\,
w_{\alpha \alpha}({\bf t}, z)\bar w_{\alpha \alpha}({\bf t}-
[a^{-1}]_{\alpha}, z)=0
\eeq
for $\beta =\alpha$. The residue calculus applied to (\ref{m10}), (\ref{m10a})
and (\ref{m11}) yields, respectively:
\beq\label{m12}
w_{\alpha \alpha}({\bf t}, a)\bar w_{\alpha \beta}({\bf t}-[a^{-1}]_{\alpha}, a)=
w_{\alpha \beta }({\bf t}),
\eeq
\beq\label{m12a}
w_{\alpha \beta}({\bf t}, a)\bar w_{\beta \beta}({\bf t}-[a^{-1}]_{\beta}, a)=
\bar w_{\alpha \beta }({\bf t}-[a^{-1}]_{\beta}),
\eeq
\beq\label{m13}
w_{\alpha \alpha}({\bf t}, a)
\bar w_{\alpha \alpha}({\bf t}-[a^{-1}]_{\alpha}, a)=1.
\eeq
Tending $a\to \infty$ in (\ref{m12}), (\ref{m12a}), we get
\beq\label{m14}
\bar w_{\alpha \beta}({\bf t})=w_{\alpha \beta}({\bf t}).
\eeq
Using (\ref{m13}), we see from (\ref{m12}), (\ref{m12a}) that
\beq\label{m15}
\begin{array}{l}
w_{\alpha \beta}({\bf t}, a)=w_{\alpha \beta}({\bf t}-[a^{-1}]_{\beta})
w_{\beta \beta}({\bf t}, a),
\\ \\
\displaystyle{\bar w_{\alpha \beta}({\bf t}, a)=
\frac{w_{\alpha \beta}({\bf t}+
[a^{-1}]_{\alpha})}{w_{\alpha \alpha}({\bf t}+[a^{-1}]_{\alpha}, a)}}.
\end{array}
\eeq
Next, we set ${\bf t}-{\bf t}'=[a^{-1}]_{\alpha}+[b^{-1}]_{\alpha}$ in (\ref{m7}). 
At $\beta =\alpha$ the residue calculus yields
$$
w_{\alpha \alpha}({\bf t}, a)\bar w_{\alpha \alpha}({\bf t}-
[a^{-1}]_{\alpha}-[b^{-1}]_{\alpha}, a)=
w_{\alpha \alpha}({\bf t}, b)\bar w_{\alpha \alpha}({\bf t}-
[a^{-1}]_{\alpha}-[b^{-1}]_{\alpha}, b)
$$
Taking into account (\ref{m13}), we can write this condition as
\beq\label{m16}
\frac{w_{\alpha \alpha}({\bf t}, a)}{w_{\alpha \alpha}
({\bf t}-[b^{-1}]_{\alpha}, a)}=
\frac{w_{\alpha \alpha}({\bf t}, b)}{w_{\alpha \alpha}
({\bf t}-[a^{-1}]_{\alpha}, b)}.
\eeq
As is proven in \cite{DJKM83}, it follows from this condition that 
there exists a function $\tau_{\alpha \alpha}({\bf t})$ such that
\beq\label{m17}
w_{\alpha \alpha}({\bf t}, a)=\frac{\tau_{\alpha \alpha}
({\bf t}-[a^{-1}]_{\alpha})}{\tau_{\alpha \alpha}({\bf t})}.
\eeq
Finally, we 
set ${\bf t}-{\bf t}'=[a^{-1}]_{\alpha}+[b^{-1}]_{\beta}$ in (\ref{m7}).
The residue calculus yields
$$
w_{\alpha \alpha}({\bf t}, a)\bar w_{\alpha \beta}({\bf t}-
[a^{-1}]_{\alpha}-[b^{-1}]_{\beta}, a)=
w_{\alpha \beta}({\bf t}, b)\bar w_{\beta \beta}({\bf t}-
[a^{-1}]_{\alpha}-[b^{-1}]_{\beta}, b).
$$
Substitute here
$$
\bar w_{\alpha \beta}({\bf t}-[a^{-1}]_{\alpha}-[b^{-1}]_{\beta}, a)=
\frac{w_{\alpha \beta}({\bf t}-b^{-1}]_{\beta})}{
w_{\alpha \alpha}({\bf t}-b^{-1}]_{\beta},a)}
$$
as in the second equation in (\ref{m15}),
$$
w_{\alpha \beta}({\bf t}, b)=w_{\alpha \beta}(({\bf t}-b^{-1}]_{\beta})
w_{\beta \beta}({\bf t}, b)
$$
as in the first equation in (\ref{m15}),
$$
\bar w_{\beta \beta}({\bf t}-
[a^{-1}]_{\alpha}-[b^{-1}]_{\beta}, b)=
\frac{1}{w_{\beta \beta}({\bf t}-
[a^{-1}]_{\alpha}, b)}
$$
as in (\ref{m13}) with the change $\alpha \to \beta$. Then 
$w_{\alpha \beta}({\bf t}-[b^{-1}]_{\beta})$ cancels and we
obtain the relation
\beq\label{m18}
\frac{w_{\alpha \alpha}({\bf t}, a)}{w_{\alpha \alpha }
({\bf t}-[b^{-1}]_{\beta}, a)}=
\frac{w_{\beta \beta}({\bf t}, b)}{w_{\beta \beta }
({\bf t}-[a^{-1}]_{\alpha}, b)}.
\eeq
Note that at $\beta =\alpha$ we get (\ref{m16}). 
As is proven in \cite{D97}, the condition (\ref{m18}) implies that
$w_{\alpha \alpha}({\bf t}, a)$ actually does not depend on the 
index $\alpha$ and thus $\tau_{\alpha \alpha}$ in (\ref{m17}) 
is the same for all $\alpha$: 
$\tau_{\alpha \alpha}({\bf t})=\tau ({\bf t})$.

Let us present the argument from \cite{D97}. 
Denote $f_{\alpha}({\bf t}, z)=
\log w_{\alpha \alpha}({\bf t}, z)$, then (\ref{m18}) acquires the form
\beq\label{m18a}
f_{\alpha}({\bf t}-[b^{-1}]_{\beta}, a)-f_{\alpha}({\bf t}, a)
=f_{\beta}({\bf t}-[a^{-1}]_{\alpha}, b)-f_{\beta}({\bf t}, b).
\eeq
To save the space, we introduce the differential operator
$$
\p_{\alpha}(z)=\p_z-\sum_{k\geq 0}z^{-k-1}\p_{t_{\alpha , k}}
$$
and apply $\p_{\alpha}(a)$ to the both sides of (\ref{m18a}).
We get
$$
\p_{\alpha}(a)f_{\alpha}({\bf t}-[b^{-1}]_{\beta}, a)-
\p_{\alpha}(a)f_{\beta}({\bf t}, b)
=\sum_{k\geq 0}a^{-k-1}\p_{t_{\alpha , k}}f_{\beta}({\bf t}, b).
$$
Multiply both sides of this equality by $a^i$ and take the residue:
\beq\label{m18b}
r_{\alpha, i}({\bf t})=r_{\alpha, i}({\bf t}-[b^{-1}]_{\beta})-
\p_{t_{\alpha , i}}f_{\beta}({\bf t}, b),
\eeq
where $r_{\alpha , i}({\bf t})=\mbox{res}_z (z^i \p_{\alpha} (z)
f_{\alpha}({\bf t}, z))$ and the residue is defined as coefficient 
at $z^{-1}$. Writing the same equality with indices $\gamma , j$ 
instead of $\alpha , i$, applying $\p_{t_{\gamma, j}}$ to the former
and $\p_{t_{\alpha, i}}$ to the latter and subtracting one from the 
other, we get:
$$
\p_{t_{\gamma, j}}r_{\alpha , i}({\bf t}-[b^{-1}]_{\beta})-
\p_{t_{\alpha, i}}r_{\gamma , j}({\bf t}-[b^{-1}]_{\beta})=
\p_{t_{\gamma, j}}r_{\alpha , i}({\bf t})-
\p_{t_{\alpha, i}}r_{\gamma , j}({\bf t}).
$$
This means that $\p_{t_{\gamma, j}}r_{\alpha , i}({\bf t})-
\p_{t_{\alpha, i}}r_{\gamma , j}({\bf t})=\mbox{const}$. From the 
definition of $r_{\alpha , i}$ it follows that the constant is zero. 
Therefore,
$$
\p_{t_{\gamma, j}}r_{\alpha , i}({\bf t})=
\p_{t_{\alpha, i}}r_{\gamma , j}({\bf t})
$$
which implies the existence of a function $\tau ({\bf t})$ such that
$r_{\alpha , i}({\bf t})=\p_{t_{\alpha , i}}\log \tau ({\bf t})$.
From (\ref{m18b}) we then see that
$$
\p_{t_{\alpha , i}}f_{\beta}({\bf t}, b)=
\p_{t_{\alpha , i}}\Bigl (\log \tau ({\bf t}-[b^{-1}]_{\beta})-
\log \tau ({\bf t})\Bigr ).
$$
Integrating, we get
$$
w_{\alpha \alpha}({\bf t}, z)=c(z)
\frac{\tau ({\bf t}-[z^{-1}]_{\alpha})}{\tau ({\bf t})},
$$
where the constant $c(z)$ can be eliminated by multiplying the 
tau-function by exponent of a linear form in the times.
Therefore,
one can write (\ref{m17}) in the form
\beq\label{m20}
w_{\alpha \alpha}({\bf t}, a)=\frac{\tau
({\bf t}-[a^{-1}]_{\alpha})}{\tau ({\bf t})},
\eeq
where the function $\tau ({\bf t})$ already does not depend on 
the index $\alpha$. Plugging (\ref{m20}) into (\ref{m15}), we get
the equations
\beq\label{m21}
\begin{array}{l}
\displaystyle{
w_{\alpha \beta}({\bf t}, a)=w_{\alpha \beta}({\bf t}-[a^{-1}]_{\beta})
\, \frac{\tau ({\bf t}-[a^{-1}]_{\beta})}{\tau ({\bf t})},}
\\ \\
\displaystyle{
\bar w_{\alpha \beta}({\bf t}, a)=w_{\alpha \beta}({\bf t}+[a^{-1}]_{\alpha})
\, \frac{\tau ({\bf t}+[a^{-1}]_{\alpha})}{\tau ({\bf t})}}
\end{array}
\eeq
which are of the form (\ref{m2}) with
$\tau_{\alpha \beta}({\bf t})=w_{\alpha \beta}({\bf t})\tau ({\bf t})$. 

\section{The multi-component CKP hierarchy}

The multi-component CKP hierarchy can be regarded as a subhierarchy
of the KP one. 

\subsection{The Lax-Sato formalism}

The set of independent variables (``times'') is
\beq\label{multi1}
{\bf t}^{\rm o}=\{{\bf t}_1^{\rm o}, {\bf t}_2^{\rm o}, \ldots , 
{\bf t}_n^{\rm o}\}, \quad
{\bf t}_{\alpha}^{\rm o}=\{t_{\alpha , 1}, t_{\alpha , 3}, t_{\alpha , 5},
\ldots \}, \quad \alpha =1, \ldots , n.
\eeq
They are divided into $n$ infinite groups, and the variables in each group are
indexed by positive odd numbers.
We also introduce the variable $x$ according to (\ref{multi2}).

The main object is the Lax operator which is a pseudo-differential operator of the form
\beq\label{multi3a}
{\cal L}=\p_x +u_1\p_x^{-1}+u_2 \p_x^{-2}+\ldots
\eeq
where the coefficients $u_i=u_i(x)$ are $n\! \times \! n$ matrices
with the constraint
\beq\label{multi4a}
{\cal L}^{\dag}=-{\cal L}.
\eeq
Here $\dag$ means the formal adjoint
defined by the rule $\Bigl (f(x)\circ 
\p_x^{n}\Bigr )^{\dag}=(-\p_x)^n \circ f^{\dag}(x)$
and $f^{\dag}$ is the transposed matrix $f$.
Besides, there are
other matrix pseudo-differential operators ${\cal R}_1, \ldots , {\cal R}_n$
of the form
\beq\label{multi5a}
{\cal R}_{\alpha}=E_{\alpha}+ u_{\alpha , 1}\p_x^{-1}+u_{\alpha , 2}
\p_x^{-2}+\ldots ,
\eeq
where $E_{\alpha}$ is the $n \! \times \! n$ matrix with the $(\alpha , \alpha )$ element
equal to 1 and all other components equal to $0$, and $u_{\alpha , i}$ are also
$n \! \times \! n$ matrices. The operators ${\cal L}$, ${\cal R}_1, 
\ldots , {\cal R}_n$
satisfy the same conditions (\ref{multi6}) 
as in the multi-component KP hierarchy.
In the CKP case the operators ${\cal R}_{\alpha}$ satisfy the constraint
\beq\label{multi7a}
{\cal R}^{\dag}_{\alpha}={\cal R}_{\alpha}.
\eeq
The Lax equations of the hierarchy which define evolution in the times read
\beq\label{multi8a}
\p_{t_{\alpha , k}}{\cal L}=[B_{\alpha , k}, \, {\cal L}], \quad
\p_{t_{\alpha , k}}{\cal R}_{\beta}=[B_{\alpha , k}, \, {\cal R}_{\beta}],
\quad B_{\alpha , k} = \Bigl ({\cal L}^k{\cal R}_{\alpha}\Bigr )_+,
\quad k=1,3,5, \ldots .
\eeq
Let us check that the constraints (\ref{multi4a}), (\ref{multi7a}) 
are preserved
by the evolution. Indeed, we have
\beq\label{multi9}
\p_{t_{\alpha , k}}({\cal L}+{\cal L}^{\dag})=[B_{\alpha , k}, \, {\cal L}]-
[B^{\dag}_{\alpha , k}, \, {\cal L}^{\dag}]=0
\eeq
and
\beq\label{multi10}
\p_{t_{\alpha , k}}({\cal R}_{\beta}-{\cal R}_{\beta}^{\dag})=
[B_{\alpha , k}, \, {\cal R}_{\beta}]+[B^{\dag}_{\alpha , k}, \, {\cal R}^{\dag}_{\beta}]=0
\eeq
since $B_{\alpha, k}^{\dag}=-B_{\alpha, k}$ for odd $k$.

Let ${\cal W}$ be the wave operator (\ref{m113})
defined by (\ref{m113}) with ${\bf t}^{\rm o}$ instead of 
${\bf t}$, which satisfies 
(\ref{multi12}).
The constraints (\ref{multi4a}), (\ref{multi7a}) imply that
${\cal W}^{\dag}{\cal W}$ commutes with $\p_x$ and $E_{\alpha}$, i.e.,
it is a pseudo-differential operator with constant coefficients commuting with $E_{\alpha}$.
The ambiguity in the definition of the dressing operator 
can be fixed by demanding that
${\cal W}^{\dag}{\cal W}=I$, i.e.
\beq\label{multi13a}
{\cal W}^{\dag}={\cal W}^{-1}.
\eeq

Let us introduce the matrix wave function
\beq\label{multi14}
\Psi (x,{\bf t}^{\rm o}, z)={\cal W} \exp \Bigl ( xzI +\sum_{\alpha =1}^n E_{\alpha}
\xi_{\rm o} ({\bf t}_{\alpha}^{\rm o}, z)\Bigr ),
\eeq
where 
\beq\label{multi15}
\xi_{\rm o} ({\bf t}_{\alpha}^{\rm o}, z)=
\sum_{k\geq 1, \,\, {\rm odd}}t_{\alpha , k}z^k.
\eeq
The wave function has the expansion
\beq\label{multi16}
\Psi_{\alpha \beta} (x,{\bf t}^{\rm o}, z)=\Biggl (\delta_{\alpha \beta}+
\sum_{k\geq 1} \xi_{k, \alpha \beta}(x, {\bf t}^{\rm o})z^{-k}\Biggr )
e^{\xi_{\rm o} ({\bf t}_{\beta}^{\rm o}, z)}.
\eeq
as $z\to \infty$ and satisfies the linear equations
\beq\label{multi17}
{\cal L}\Psi =z\Psi , \qquad
\p_{t_{\alpha , k}}\Psi =B_{\alpha , k}\Psi \quad (\mbox{$k$ odd}).
\eeq

The CKP constraint on the wave operator (\ref{multi13a}) implies that
\beq\label{multi18}
\Psi^{*}(x, {\bf t}^{\rm o}, z)=\Psi^{\dag}(x, {\bf t}^{\rm o}, -z),
\eeq
where $\Psi^{\dag}$ is the transposed matrix. Taking into account 
(\ref{m2}) 
which holds in the CKP case with the modification ${\bf t}$ to 
${\bf t}^{\rm o}$, 
we can represent this relation as the following conditions
for the KP tau-function:
\beq\label{multi19}
\left \{ \begin{array}{l}
\tau (x, {\bf t}^{\rm o}, -[z^{-1}]_{\alpha}^{\rm e})=
\tau (x, {\bf t}^{\rm o}, [z^{-1}]_{\alpha}^{\rm e}),
\\ \\
\tau_{\beta \alpha} (x, {\bf t}^{\rm o}, -[z^{-1}]_{\alpha}^{\rm e})=
-\tau_{\alpha \beta} (x, {\bf t}^{\rm o}, [z^{-1}]_{\alpha}^{\rm e}),
\quad \alpha \neq \beta ,
\end{array}
\right.
\eeq
where $\pm [z^{-1}]_{\alpha}^{\rm e}$ means that the ``even'' times 
$t_{\alpha , 2k}$ are equal to $\pm \frac{1}{2k}\, z^{-2k}$. Expanding 
(\ref{multi19}) in powers of $z^{-1}$ as $z\to \infty$, 
in the leading order we obtain the necessary
conditions 
\beq\label{multi20}
\left \{ \begin{array}{l}
\p_{t_{\alpha ,2}}\tau (x, {\bf t}^{\rm o})\Bigr |_{{\bf t}^{\rm e}=0}=0,
\\ \\
\tau_{\beta \alpha} (x, {\bf t}^{\rm o})\Bigr |_{{\bf t}^{\rm e}=0}=
-\tau_{\alpha \beta} (x, {\bf t}^{\rm o})\Bigr |_{{\bf t}^{\rm e}=0},
\quad \alpha \neq \beta ,
\end{array}
\right.
\eeq
where ${\bf t}^{\rm e}$ is the set of ``even'' times. 

\subsection{Existence of the tau-function $\tau^{\mbox{\tiny CKP}}$}

In this section we put $x=0$ without loss of generality and will not
indicate the dependence of $x$ explicitly.

The bilinear relation (\ref{m3}) for the CKP hierarchy acquires the form
\beq\label{mtau1}
\sum_{\gamma}\oint_{C_{\infty}} dz \Psi_{\alpha \gamma}({\bf t}^{\rm o},
z)\Psi_{\beta \gamma}({{\bf t}^{\rm o}}{'},-z)=0.
\eeq
This relation is as a starting point for the proof of
existence of the tau-function $\tau^{\rm CKP}
({\bf t}^{\rm o})$ of the CKP hierarchy which depends 
on ``odd'' times only. 

Let us represent the wave function in the form
\beq\label{mtau2}
\Psi_{\alpha \beta}({\bf t}^{\rm o}, z)=
\epsilon_{\alpha \beta}\,e^{\xi_{\rm o} ({\bf t}_{\beta}^{\rm o}, z)}
z^{\delta_{\alpha \beta}-1}w_{\alpha \beta}({\bf t}^{\rm o}, z),
\eeq
with some matrix function $w({\bf t}^{\rm o}, z)$. 
Then we can write the bilinear relation as
\beq\label{mtau3}
\sum_{\gamma} \epsilon_{\alpha \gamma}\epsilon_{\beta \gamma}
(-1)^{\delta_{\beta \gamma}-1}
\oint_{C_{\infty}}\! dz \, 
e^{\xi_{\rm o} ({\bf t}_{\gamma}^{\rm o}-{{\bf t}_{\gamma}^{\rm o}}', z)}
z^{\delta_{\alpha \gamma}+\delta_{\beta \gamma}-2}
w_{\alpha \gamma}({\bf t}^{\rm o}, z)
w_{\beta \gamma}({{\bf t}^{\rm o}}', -z)=0.
\eeq
In the calculations below, we need the values
\beq\label{mtau4}
w_{\alpha \beta}({\bf t}^{\rm o}, \infty )
=:w_{\alpha \beta}({\bf t}^{\rm o}).
\eeq
One can always normalize them as
$
w_{\alpha \alpha}({\bf t}^{\rm o})=1.
$

We first consider the case ${\bf t}_{\gamma}^{\rm o}=
{{\bf t}_{\gamma}^{\rm o}}'$ for all $\gamma$ except $\gamma =\alpha$.
Then the bilinear relation (\ref{mtau3}) at $\beta =\alpha$ reads
\beq\label{mtau5}
\oint_{C_{\infty}}\! dz \, 
e^{\xi_{\rm o} ({\bf t}_{\alpha}^{\rm o}-{{\bf t}_{\alpha}^{\rm o}}', z)}
w_{\alpha \alpha}({\bf t}^{\rm o}, z)
w_{\alpha \alpha}({{\bf t}^{\rm o}}', -z)=0.
\eeq
For each $\alpha$, it has the same form as the bilinear relation
for the one-component CKP hierarchy. 
As is proven in \cite{KZ21}, 
it follows from (\ref{mtau5}) that there exist functions 
$\tau^{\mbox{\tiny CKP}}_{\alpha \alpha}({\bf t}^{\rm o})$ such that
\beq\label{mtau6}
w_{\alpha \alpha}({\bf t}^{\rm o}, z)=f_{\alpha}^{1/2}({\bf t}^{\rm o}, z)
\frac{\tau^{\mbox{\tiny CKP}}_{\alpha \alpha}({\bf t}^{\rm o}-
2[z^{-1}]_{\alpha}^{\rm o})}{\tau^{\mbox{\tiny CKP}}_{\alpha \alpha}({\bf t}^{\rm o})},
\eeq
where
$$
\left ({\bf t}\pm [z^{-1}]_{\gamma}^{\rm o}
\right )_{\alpha ,k}=t_{\alpha , k}\pm
\delta_{\alpha \gamma} \frac{z^{-k}}{k}, \quad \mbox{$k$ odd},
$$
and
\beq\label{mtau7}
f_{\alpha}({\bf t}^{\rm o}, z)=1+z^{-1}\p_{t_{\alpha ,1}}
\log \frac{\tau^{\mbox{\tiny CKP}}_{\alpha \alpha}({\bf t}^{\rm o}-
2[z^{-1}]_{\alpha}^{\rm o})}{\tau^{\mbox{\tiny CKP}}_{\alpha \alpha}({\bf t}^{\rm o})}.
\eeq
Note that $f_{\alpha}({\bf t}^{\rm o}, z)=f_{\alpha}
({\bf t}^{\rm o}-2[z^{-1}]^{\rm o}_{\alpha}, -z)$.
In particular, we can put
${\bf t}^{\rm o}-{{\bf t}^{\rm o}}'=2[a^{-1}]_{\alpha}^{\rm o}$, 
then
$\displaystyle{
e^{\xi_{\rm o} ({\bf t}_{\alpha}^{\rm o}-{{\bf t}_{\alpha}^{\rm o}}', z)}=
\frac{a+z}{a-z}}$ and the residue calculus yields
\beq\label{mtau8}
w_{\alpha \alpha}({\bf t}^{\rm o}, a)
w_{\alpha \alpha}({\bf t}^{\rm o}-2[a^{-1}]^{\rm o}_{\alpha}, -a)=
f_{\alpha}({\bf t}^{\rm o}, a).
\eeq
The left hand side of (\ref{mtau8}) comes from the residue at $z=a$,
the right hand side is the residue at infinity. In fact (\ref{mtau8})
directly follows from (\ref{mtau6}).

Next, we put
${\bf t}^{\rm o}-{{\bf t}^{\rm o}}'=2[a^{-1}]_{\alpha}^{\rm o}$,
${\bf t}^{\rm o}-{{\bf t}^{\rm o}}'=2[a^{-1}]_{\beta}^{\rm o}$
and plug this into (\ref{mtau3}). We have, respectively, 
for $\beta \neq \alpha$:
$$
\oint_{C_{\infty}}\frac{dz}{z}\, \frac{a+z}{a-z} \,
w_{\alpha \alpha}({\bf t}^{\rm o}, z)
w_{\beta \alpha}({\bf t}^{\rm o}-2[a^{-1}]^{\rm o}_{\alpha}, -z)
$$
$$
+\oint_{C_{\infty}}\frac{dz}{z}\,
w_{\alpha \beta}({\bf t}^{\rm o}, z)
w_{\beta \beta}({\bf t}^{\rm o}-2[a^{-1}]^{\rm o}_{\alpha}, -z)=0,
$$
$$
\oint_{C_{\infty}}\frac{dz}{z}\, \frac{a+z}{a-z} \,
w_{\alpha \beta}({\bf t}^{\rm o}, z)
w_{\beta \beta}({\bf t}^{\rm o}-2[a^{-1}]^{\rm o}_{\beta}, -z)
$$
$$
+\oint_{C_{\infty}}\frac{dz}{z}\,
w_{\alpha \alpha}({\bf t}^{\rm o}, z)
w_{\beta \alpha}({\bf t}^{\rm o}-2[a^{-1}]^{\rm o}_{\beta}, -z)=0.
$$
The residue calculus yields
\beq\label{mtau9}
2w_{\alpha \alpha}({\bf t}^{\rm o}, a)
w_{\beta \alpha}({\bf t}^{\rm o}-2[a^{-1}]^{\rm o}_{\alpha}, -a)=
w_{\beta \alpha}({\bf t}^{\rm o}-2[a^{-1}]^{\rm o}_{\alpha})-
w_{\alpha \beta}({\bf t}^{\rm o}),
\eeq
\beq\label{mtau9a}
2w_{\alpha \beta}({\bf t}^{\rm o}, a)
w_{\beta \beta}({\bf t}^{\rm o}-2[a^{-1}]^{\rm o}_{\beta}, -a)=
w_{\alpha \beta}({\bf t}^{\rm o})-
w_{\beta \alpha}({\bf t}^{\rm o}-2[a^{-1}]^{\rm o}_{\beta}).
\eeq
Tending here $a$ to $\infty$, we see that
\beq\label{mtau10}
w_{\beta \alpha }({\bf t}^{\rm o})=-w_{\alpha \beta}({\bf t}^{\rm o}),
\quad \alpha \neq \beta 
\eeq
(cf. the second equation in (\ref{multi20})).
From these relations we conclude that
\beq\label{mtau11}
w_{\alpha \beta}({\bf t}^{\rm o}, a)=
\frac{w_{\alpha \beta}({\bf t}^{\rm o})+
w_{\alpha \beta}({\bf t}^{\rm o}-2[a^{-1}]^{\rm o}_{\beta})}{2
w_{\beta \beta}({\bf t}^{\rm o}-2[a^{-1}]^{\rm o}_{\beta}, -a)},
\eeq
or, using (\ref{mtau6}),
\beq\label{mtau12}
w_{\alpha \beta}({\bf t}^{\rm o}, a)=
\frac{w_{\alpha \beta}({\bf t}^{\rm o})+
w_{\alpha \beta}({\bf t}^{\rm o}-2[a^{-1}]^{\rm o}_{\beta})}{2
f_{\beta}^{1/2}({\bf t}^{\rm o}, a)}\,
\frac{\tau^{\mbox{\tiny CKP}}_{\beta \beta}({\bf t}^{\rm o}-
2[a^{-1}]^{\rm o}_{\beta})}{\tau^{\mbox{\tiny CKP}}_{\beta \beta}
({\bf t}^{\rm o})}.
\eeq

Our next goal is to show that the functions $\tau^{\mbox{\tiny CKP}}_{\alpha \alpha}$
in (\ref{mtau6}) can be always made the same for all $\alpha$, i.e.,
$\tau^{\mbox{\tiny CKP}}_{\alpha \alpha}({\bf t}^{\rm o})=
\tau^{\mbox{\tiny CKP}}({\bf t}^{\rm o})$. To this end, we 
consider (\ref{mtau3}) at 
${\bf t}^{\rm o}-{{\bf t}^{\rm o}}'=2[a^{-1}]_{\alpha}^{\rm o}+
2[b^{-1}]_{\beta}^{\rm o}$ ($\alpha \neq \beta$):
$$
\oint_{C_{\infty}}\frac{dz}{z}\, \frac{a+z}{a-z} \,
w_{\alpha \alpha}({\bf t}^{\rm o}, z)
w_{\beta \alpha}({\bf t}^{\rm o}-2[a^{-1}]^{\rm o}_{\alpha}
-2[b^{-1}]^{\rm o}_{\beta}, -z)
$$
$$
+\oint_{C_{\infty}}\frac{dz}{z}\, \frac{b+z}{b-z} \,
w_{\alpha \beta}({\bf t}^{\rm o}, z)
w_{\beta \beta}({\bf t}^{\rm o}-2[a^{-1}]^{\rm o}_{\alpha}
-2[b^{-1}]^{\rm o}_{\beta}, -z)=0.
$$
The residue calculus yields:
\beq\label{mtau13}
\begin{array}{c}
2w_{\alpha \alpha}({\bf t}^{\rm o}, a)
w_{\beta \alpha}({\bf t}^{\rm o}\! -\! 2[a^{-1}]^{\rm o}_{\alpha}
\! -\! 2[b^{-1}]^{\rm o}_{\beta}, -a)\! +\! 
2w_{\alpha \beta}({\bf t}^{\rm o}, b)
w_{\beta \beta}({\bf t}^{\rm o}\! -\! 2[a^{-1}]^{\rm o}_{\alpha}
\! -\! 2[b^{-1}]^{\rm o}_{\beta}, -b)
\\ \\
=w_{\alpha \beta}({\bf t}^{\rm o})
-w_{\alpha \beta}({\bf t}^{\rm o}-2[a^{-1}]^{\rm o}_{\alpha}
-2[b^{-1}]^{\rm o}_{\beta}).
\end{array}
\eeq
Plugging here (\ref{mtau11}) and shifting ${\bf t}^{\rm o}\to
{\bf t}^{\rm o}+2[b^{-1}]^{\rm o}_{\beta}$ with the subsequent 
change of sign $b\to -b$, we get the equation
\beq\label{mtau14}
\begin{array}{l}
\displaystyle{\frac{w_{\beta \beta}({\bf t}^{\rm o}-
2[a^{-1}]^{\rm o}_{\alpha}, b)}{w_{\beta \beta}({\bf t}^{\rm o}, b)}
\Bigl (w_{\alpha \beta}({\bf t}^{\rm o})+w_{\alpha \beta}({\bf t}^{\rm o}
-2[b^{-1}]^{\rm o}_{\beta})\Bigr )}
\\ \\
\phantom{aaaaaaaaaaaaaaaa}-
\displaystyle{\frac{w_{\alpha \alpha}({\bf t}^{\rm o}-
2[b^{-1}]^{\rm o}_{\beta}, a)}{w_{\alpha \alpha}({\bf t}^{\rm o}, a)}
\Bigl (w_{\alpha \beta}({\bf t}^{\rm o})+w_{\alpha \beta}({\bf t}^{\rm o}
-2[a^{-1}]^{\rm o}_{\alpha})\Bigr )}
\\ \\
\phantom{aaaaaaaaaaa}=w_{\alpha \beta}({\bf t}^{\rm o}-
2[b^{-1}]^{\rm o}_{\beta})-w_{\alpha \beta}({\bf t}^{\rm o}-
2[a^{-1}]^{\rm o}_{\alpha}).
\end{array}
\eeq
Shifting ${\bf t}^{\rm o}\to
{\bf t}^{\rm o}+2[a^{-1}]^{\rm o}_{\alpha}+2[b^{-1}]^{\rm o}_{\beta}$ with the subsequent 
change of signs $a\to -a$, $b\to -b$ and using (\ref{mtau8}), 
we obtain another equation:
\beq\label{mtau15}
\begin{array}{l}
\displaystyle{\frac{f_{\beta}({\bf t}^{\rm o}, b)}{f_{\beta}({\bf t}^{\rm o}
\! -\! 2[a^{-1}]^{\rm o}_{\alpha}, b)}
\frac{w_{\beta \beta}({\bf t}^{\rm o}\! -\!
2[a^{-1}]^{\rm o}_{\alpha}, b)}{w_{\beta \beta}({\bf t}^{\rm o}, b)}
\Bigl (w_{\alpha \beta}({\bf t}^{\rm o}\! -\! 2[a^{-1}]^{\rm o}_{\alpha}
\! - \!2[b^{-1}]^{\rm o}_{\beta})\! +\! w_{\alpha \beta}({\bf t}^{\rm o}
\! -\! 2[a^{-1}]^{\rm o}_{\alpha})\Bigr )}
\\ \\
\phantom{}-
\displaystyle{\frac{f_{\alpha}({\bf t}^{\rm o}, a)}{f_{\alpha}({\bf t}^{\rm o}
\! -\! 2[b^{-1}]^{\rm o}_{\beta}, a)}
\frac{w_{\alpha \alpha}({\bf t}^{\rm o}\! -\! 
2[b^{-1}]^{\rm o}_{\beta}, a)}{w_{\alpha \alpha}({\bf t}^{\rm o}, a)}
\Bigl (w_{\alpha \beta}({\bf t}^{\rm o}\! -\! 2[a^{-1}]^{\rm o}_{\alpha}
\! -\! 2[b^{-1}]^{\rm o}_{\beta})\! +\! w_{\alpha \beta}({\bf t}^{\rm o}
\! -\! 2[b^{-1}]^{\rm o}_{\beta})\Bigr )}
\\ \\
\phantom{aaaaaaaaaaa}=w_{\alpha \beta}({\bf t}^{\rm o}-
2[a^{-1}]^{\rm o}_{\alpha})-w_{\alpha \beta}({\bf t}^{\rm o}-
2[b^{-1}]^{\rm o}_{\beta}).
\end{array}
\eeq
Using the representation (\ref{mtau6}), we can rewrite equations 
(\ref{mtau14}), (\ref{mtau15}) as a system of two linear equations
for two ``variables''
\beq\label{mtau16}
\begin{array}{l}
\displaystyle{
X_{\alpha}=\frac{\tau^{\mbox{\tiny CKP}}_{\alpha \alpha}({\bf t}^{\rm o})
\tau^{\mbox{\tiny CKP}}_{\alpha \alpha}({\bf t}^{\rm o}- 2[a^{-1}]^{\rm o}_{\alpha}
- 2[b^{-1}]^{\rm o}_{\beta})}{\tau^{\mbox{\tiny CKP}}_{\alpha \alpha}({\bf t}^{\rm o}
- 2[a^{-1}]^{\rm o}_{\alpha})\tau^{\mbox{\tiny CKP}}_{\alpha \alpha}({\bf t}^{\rm o}
- 2[b^{-1}]^{\rm o}_{\beta})}}
\\ \\
\displaystyle{
X_{\beta}=\frac{\tau^{\mbox{\tiny CKP}}_{\beta \beta}({\bf t}^{\rm o})
\tau^{\mbox{\tiny CKP}}_{\beta \beta}({\bf t}^{\rm o}- 2[a^{-1}]^{\rm o}_{\alpha}
- 2[b^{-1}]^{\rm o}_{\beta})}{\tau^{\mbox{\tiny CKP}}_{\beta \beta}({\bf t}^{\rm o}
- 2[a^{-1}]^{\rm o}_{\alpha})\tau^{\mbox{\tiny CKP}}_{\beta \beta}({\bf t}^{\rm o}
- 2[b^{-1}]^{\rm o}_{\beta})}}.
\end{array}
\eeq
In order to write the system in a compact form, it is convenient to use
the following short-hand notation: for any function $F({\bf t}^{\rm o})$
we will write 
$$
F({\bf t}^{\rm o}- 2[a^{-1}]^{\rm o}_{\alpha})=F^{[a]}, \quad
F({\bf t}^{\rm o}- 2[b^{-1}]^{\rm o}_{\beta})=F^{[b]},
$$
$$
F({\bf t}^{\rm o}- 2[a^{-1}]^{\rm o}_{\alpha}- 
2[b^{-1}]^{\rm o}_{\beta})=F^{[ab]}
$$
(for example, $f_{\alpha}^{[b]}(a)=f_{\alpha}
({\bf t}^{\rm o}- 2[b^{-1}]^{\rm o}_{\beta}, a)$).
In this notation, equations (\ref{mtau14}), (\ref{mtau15}) read:
\beq\label{mtau17}
\left (\frac{f_{\beta}^{[a]}(b)}{f_{\beta}(b)}\right )^{1/2}\!\!
\Bigl (w_{\alpha \beta}+w_{\alpha \beta}^{[b]}\Bigr )X_{\beta} -
\left (\frac{f_{\alpha}^{[b]}(a)}{f_{\alpha}(a)}\right )^{1/2}\!\!
\Bigl (w_{\alpha \beta}+w_{\alpha \beta}^{[a]}\Bigr )X_{\alpha}
=w_{\alpha \beta}^{[b]}-w_{\alpha \beta}^{[a]},
\eeq
\beq\label{mtau18}
\left (\frac{f_{\beta}(b)}{f^{[a]}_{\beta}(b)}\right )^{1/2}\!\!
\Bigl (w_{\alpha \beta}^{[ab]}+w_{\alpha \beta}^{[a]}\Bigr )X_{\beta} -
\left (\frac{f_{\alpha}(a)}{f_{\alpha}^{[b]}(a)}\right )^{1/2}\!\!
\Bigl (w_{\alpha \beta}^{[ab]}+w_{\alpha \beta}^{[b]}\Bigr )X_{\alpha}
=w_{\alpha \beta}^{[a]}-w_{\alpha \beta}^{[b]}.
\eeq
Shifting ${\bf t}^{\rm o}\to
{\bf t}^{\rm o}+2[a^{-1}]^{\rm o}_{\alpha}$ with the subsequent 
change of sign $a\to -a$ (and ${\bf t}^{\rm o}\to
{\bf t}^{\rm o}+2[b^{-1}]^{\rm o}_{\beta}$ with the subsequent 
change of sign $b\to -b$), we get two more equations from (\ref{mtau18}):
\beq\label{mtau19}
\left (\frac{f_{\beta}^{[a]}(b)}{f_{\beta}(b)}\right )^{1/2}\!\!
\Bigl (w_{\alpha \beta}+w_{\alpha \beta}^{[b]}\Bigr )X_{\beta}^{-1} -
\left (\frac{f_{\alpha}^{}(a)}{f_{\alpha}^{[b]}(a)}\right )^{1/2}\!\!
\Bigl (w_{\alpha \beta}^{[b]}+w_{\alpha \beta}^{[ab]}\Bigr )X_{\alpha}^{-1}
=w_{\alpha \beta}^{}-w_{\alpha \beta}^{[ab]},
\eeq
\beq\label{mtau20}
\left (\frac{f_{\beta}^{}(b)}{f_{\beta}^{[a]}(b)}\right )^{1/2}\!\!
\Bigl (w_{\alpha \beta}^{[a]}+w_{\alpha \beta}^{[ab]}\Bigr )X_{\beta}^{-1} -
\left (\frac{f_{\alpha}^{[b]}(a)}{f_{\alpha}(a)}\right )^{1/2}\!\!
\Bigl (w_{\alpha \beta}+w_{\alpha \beta}^{[a]}\Bigr )X_{\alpha}^{-1}
=w_{\alpha \beta}^{[ab]}-w_{\alpha \beta}^{}.
\eeq
The next step is to multiply the both sides of equations
(\ref{mtau17}), (\ref{mtau18}) by the corresponding sides of equations
(\ref{mtau19}), (\ref{mtau20}). Performing this procedure with equations
(\ref{mtau17}), (\ref{mtau19}) and then with
(\ref{mtau18}), (\ref{mtau20}), we get a system of linear
equations for $X_{\beta}/X_{\alpha}$ and $X_{\alpha}/X_{\beta}$ of the form
$$
\left \{ 
\begin{array}{l}
\displaystyle{c \,\frac{X_{\beta}}{X_{\alpha}}+d \,\frac{X_{\alpha}}{X_{\beta}}=h}
\\ \\
\displaystyle{
d \, \frac{X_{\beta}}{X_{\alpha}}+c \,\frac{X_{\alpha}}{X_{\beta}}=h}
\end{array}
\right.
$$
with
$$
c=\left (\frac{f_{\alpha}^{[b]}(a)
f_{\beta}(b)}{f_{\alpha}(a)f_{\beta}^{[a]}(b)}\right )^{1/2} 
\!\! \Bigl (w_{\alpha \beta}+w_{\alpha \beta}^{[a]}\Bigr ),
\quad
d=\left (\frac{f_{\alpha}^{[b]}(a)
f_{\beta}^{[a]}(b)}{f_{\alpha}(a)f_{\beta}(b)}\right )^{-1/2}
\!\! \Bigl (w_{\alpha \beta}^{ab}+w_{\alpha \beta}^{[b]}\Bigr ),
$$
$$
h=w_{\alpha \beta}+w_{\alpha \beta}^{[b]}+
\frac{f_{\beta}(b)}{f_{\beta}^{[a]}(b)}(w_{\alpha \beta}^{[b]}+
w_{\alpha \beta}^{[ab]}).
$$
A similar system is obtained in the same way from equations
(\ref{mtau18}), (\ref{mtau19}) and
(\ref{mtau18}), (\ref{mtau20}).
It immediately follows from this system that 
$(X_{\alpha}/X_{\beta})^2=1$, so $X_{\alpha}/X_{\beta}=\pm 1$. 
Since at $b^{-1}=0$ we have 
$X_{\alpha}=X_{\beta}=1$, and the function
$X_{\alpha}/X_{\beta}$ is continuous in $b^{-1}$, 
we conclude that $X_{\alpha}=X_{\beta}$ for all $a,b$.
Then the same argument as before allows one to conclude that
the function $\tau^{\mbox{\tiny CKP}}_{\alpha \alpha}$ can be chosen to be 
one and the same for all $\alpha$:
$\tau^{\mbox{\tiny CKP}}_{\alpha \alpha}({\bf t}^{\rm o})=
\tau^{\mbox{\tiny CKP}}_{}({\bf t}^{\rm o})$. 

For $\alpha \neq \beta$, we can introduce functions 
$\tau_{\alpha \beta}^{\mbox{\tiny CKP}}({\bf t}^{\rm o})$ such that
\beq\label{mtau21}
w_{\alpha \beta}({\bf t}^{\rm o})=
\frac{\tau_{\alpha \beta}^{\mbox{\tiny CKP}}
({\bf t}^{\rm o})}{\tau^{\mbox{\tiny CKP}}({\bf t}^{\rm o})}.
\eeq
The $\tau_{\alpha \beta}^{\mbox{\tiny CKP}}$ are off-diagonal components 
of the multi-component CKP tau-function.

Finally, we present the expressions for $w_{\alpha \beta}$ 
and $w_{\alpha \alpha}$ through the tau-function:
\beq\label{mtau22}
w_{\alpha \beta}({\bf t}^{\rm o}, z)=
\left (\frac{\tau^{\mbox{\tiny CKP}}_{\alpha \beta}
({\bf t}^{\rm o})}{\tau^{\mbox{\tiny CKP}}_{}({\bf t}^{\rm o})}+
\frac{\tau^{\mbox{\tiny CKP}}_{\alpha \beta}
({\bf t}^{\rm o}-2[z^{-1}]^{\rm o}_{\beta})}{\tau^{\mbox{\tiny CKP}}_{}
({\bf t}^{\rm o}-2[z^{-1}]^{\rm o}_{\beta})}\right )
\frac{\tau^{\mbox{\tiny CKP}}_{}
({\bf t}^{\rm o}-2[z^{-1}]^{\rm o}_{\beta})}{2f_{\beta}^{1/2}(
{\bf t}^{\rm o}, z)\tau^{\mbox{\tiny CKP}}({\bf t}^{\rm o})},
\eeq
\beq\label{mtau23}
w_{\alpha \alpha}({\bf t}^{\rm o}, z)=f_{\alpha}^{1/2}({\bf t}^{\rm o}, z)
\frac{\tau^{\mbox{\tiny CKP}}_{}({\bf t}^{\rm o}-
2[z^{-1}]_{\alpha}^{\rm o})}{\tau^{\mbox{\tiny CKP}}_{}({\bf t}^{\rm o})},
\eeq
where
\beq\label{mtau24}
f_{\alpha}({\bf t}^{\rm o}, z)=1+z^{-1}\p_{t_{\alpha ,1}}
\log \frac{\tau^{\mbox{\tiny CKP}}_{}({\bf t}^{\rm o}-
2[z^{-1}]_{\alpha}^{\rm o})}{\tau^{\mbox{\tiny CKP}}_{}({\bf t}^{\rm o})}.
\eeq
Note that if we put $\beta =\alpha$ in (\ref{mtau22}), we {\it do not} get
(\ref{mtau23}). 

It remains to find how the CKP tau-function is related to the KP
tau-function $\tau ({\bf t}^{\bullet})$ satisfying the conditions
(\ref{multi20}) (by ${\bf t}^{\bullet}$ we denote the set of times
in which all ``even'' times $t_{\alpha, 2k}$ are put equal to $0$). 
Repeating the argument of \cite{KZ21}, we readily find that
\beq\label{mtau25}
\tau^{\mbox{\tiny CKP}}({\bf t}^{\rm o})=\sqrt{\vphantom{A^{A}}\tau
({\bf t}^{\bullet})}.
\eeq
For the off-diagonal components we have:
$$
w_{\alpha \beta}({\bf t}^{\rm o})=
\frac{\tau_{\alpha \beta}({\bf t}^{\bullet})}{\tau ({\bf t}^{\bullet})}=
\frac{\tau^{\mbox{\tiny CKP}}_{\alpha \beta}
({\bf t}^{\rm o})}{\tau^{\mbox{\tiny CKP}} ({\bf t}^{\rm o})},
$$
whence
\beq\label{mtau26}
\tau_{\alpha \beta}({\bf t}^{\bullet})=
\tau^{\mbox{\tiny CKP}}_{\alpha \beta}({\bf t}^{\rm o})
\tau^{\mbox{\tiny CKP}} ({\bf t}^{\rm o})
\eeq
(we have used (\ref{mtau25})). Note that (\ref{mtau26}) is valid 
for $\alpha =\beta$ as well. 

\section{Concluding remarks}

We have proved existence of the tau-function for the multi-component
CKP hierarchy and found how it is related to the multi-component 
KP tau-function. The starting point was the bilinear relation (\ref{mtau1})
for the matrix wave functions. 
The CKP tau-function $\tau^{\rm CKP}$ is a matrix 
with matrix elements $\tau^{\rm CKP}_{\alpha \beta}$ which 
are functions of the time variables
$t_{\alpha , k}$ with odd indices $k$ only. The wave functions are
expressed through the tau-function according to equations (\ref{mtau22}),
(\ref{mtau23}). Substituting these expressions into the bilinear
equation (\ref{mtau1}) for the wave functions, one can obtain an equation
for the CKP tau-function but in contrast to the KP case 
it is not bilinear. 

\section*{Acknowledgments}

\addcontentsline{toc}{section}{Acknowledgments}

This work is an output of a research project 
implemented as a part of the Basic Research Program 
at the National Research University Higher School 
of Economics (HSE University). I thank the referee for 
useful remarks and pointing
out reference \cite{D97}. 

\section*{Statements and declarations}

\subsection*{Conflict of interest} 

The author has no competing interests 
to declare that are relevant to the content of this article.

\subsection*{Data availability statement}

Data sharing is not applicable to this article as no 
datasets were generated or analysed during the current study.

\end{document}